\documentclass[aps,prb,reprint]{revtex4-1}
\usepackage[latin1]{inputenc}
\usepackage{amsmath}
\usepackage{amsfonts}
\usepackage{amssymb}
\usepackage{bm}
\usepackage{natbib}
\usepackage[dvips]{graphicx}

\begin{document}
\title{In-situ accumulated stress measurements: application to strain balanced quantum dots and quantum posts}
\date{\today}
\author{D. Alonso-\'Alvarez.}
\email{diego.alonso@imm.cnm.csic.es}
\author{B. Al\'en}
\author{J. M. Ripalda}
\author{A. Rivera}
\author{A. G. Taboada}
\author{J. M. Llorens}
\author{Y. Gonz\'alez}
\author{L. Gonz\'alez}
\author{F. Briones}
\affiliation{IMM-Instituto de Microelectr\'onica de Madrid (CNM-CSIC), Isaac Newton 8, 28760 Tres Cantos, Spain}

\pacs{codigos PAC}

\begin{abstract}
In this work we use the in-situ accumulated stress monitoring technique to evaluate the evolution of the stress during the strain balancing of InAs/GaAs quantum dots and quantum posts. The comparison of these results with simulations and other strain balanced criteria commonly used indicate that it is necessary to consider the kinematics of the process, not only the nominal values for the deposited materials. We find that the substrate temperature plays a major role on the compensation process and it is necessary to take it into account in order to achieve the optimum compensation conditions. The application of the technique to quantum posts has allowed us to fabricate nanostructures of exceptional length (120 nm). In situ accumulated measurements show that, even in shorter nanostrcutures, relaxation processes can be inhibited with the resulting increase in the material quality.
\end{abstract}
\maketitle

\section{Introduction}

In-situ characterization techniques are one of the most powerful tools to control and monitor the kinematics of the epitaxial growth of heterostructures. They can give real time information about the evolution of the growth front and the formation of nanostructures, among others. The reflection high energy electron diffraction (RHEED) system is a standard piece of equipment in most molecular beam epitaxy (MBE) reactors, and there are many works about its usage to monitor changes in surface reconstructions,~\cite{Aspnes1987, Liu1992, Tatarenko1994} the formation of quantum dots (QDs) and quantum wires (QWRs),~\cite{Lee2004} optimum conditions to fabricate them and, even, about how to get the size and shape of those nanostructures from the study of the diffracted pattern.~\cite{Lee1998} In summary it is a mature technology with a long tradition in this field. 

Despite it was introduced back in the early 90's by Schell-Sorokin and Tromp, probably less known is the in-situ accumulated stress  measurement (ASM) technique.~\cite{Schell1990} It basically consist on measuring the stress accumulated in a sample during the epitaxial growth by monitoring the changes in its curvature. The kind of information that can be extracted by this technique is very broad and ranges from the anisotropic strength of the surface reconstitutions,~\cite{Fuster2006} the study of QDs, QWRs and quantum rings formation,~\cite{Silveira2001,Fuster2004} thermal expansion in heterostructures or the formation and evolution of dislocations and plastic deformation processes.~\cite{Gonzalez2002a, Gonzalez2002b}  

In its basic form, the bending of he sample is measured using the deflection of a laser beam on the sample surface. This method is easy to implement inside a MBE reactor since all the setup is outside the vacuum chamber. In this case, a lever-shaped sample is fix from one of its ends to a special sample holder. An aperture of enough size made on the center of the holder allows the lever to bend freely. This particular requirements of the holder and the sample itself is probably what prevents a general implantation of the technique in commercial MBE reactors. 

In general, two parallel laser beams hit the sample in a direction perpendicular to the surface, one on the fix end and the other on the free one. If substrate bends, we can measure the deflection of the beam that hits the free end compared to the other beam. Using two beams reduces the noise associated with mechanical vibrations and small temperature variations. The deflection can be recorded collecting the reflected laser beams with two segmented detectors.~\cite{Fuster2006,Silveira2001,Fuster2004,Gonzalez2002a, Gonzalez2002b}

\section{Working principles}
Using the above mention geometry, the substrate curvature can then be calculated as (Stoney\'s equation):
\begin{equation}
\Delta\left(\dfrac{1}{R}\right) = \dfrac{(d-d_0)\cos \alpha}{d_H 2L}
\label{eq:curvatura}
\end{equation}
where d is the distance between the spots in the detectors, $d0$ the initial distance between spots, $d_H$ is the separation between the laser beams, $L$ the sample-detector distance, $\alpha$ is the incidence angle and $\Delta(1/R)$ is the substrate curvature variation. If the deposited layer material has a lattice parameter larger than the substrate, then it suffers a compressive stress and the substrate bends, acquiring a convex curvature ($\Delta 1/R > 0$). On the contrary, if the lattice parameter of the deposited layer is smaller than the substrate, the strain is tensile and the substrate becomes concave ($\Delta 1/R > 0$). 

Before the plastic limit, where the sample suffers from partial relief of accumulated stress through the formation of dislocations, changes in the substrate curvature and the accumulated stress can be related by means of a modified version of the Stoney's equation, to include the biaxial character of the stress in the thin layers:  
\begin{equation}
\dfrac{1}{R} = -\dfrac{6(1-\upsilon_S)\sigma h}{Y_s h_S^2}=-\dfrac{6M_S\sigma h}{h_S^2}
\label{eq:stoney}
\end{equation}
where R is the curvature radius, h is the thickness of the deposited layer, h$_S$ is the substrate thickness, $\sigma$ the stress in th elayer and M$_S$=(1-$\upsilon_S$)/Y$_S$ the biaxial modulus that relates the Young modulus (Y$_S$) and the Poisson modulus ($\upsilon_S$).This equation is valid only under the following conditions:

\begin{enumerate}
 \item The thickness of the deposited layer and the substrate are much smaller than their lateral dimensions. 
 \item The thickness of the deposited layer is much smaller than the substrate.
 \item The stress induced by the layer does not have a component in the direction normal to the sample surface.  
 \item Substrate material is linearly elastic, homogeneous and isotropic. The deposited layer must also be isotropic.
 \item Edge effects are negligible and physical properties are homogeneous in planes perpendicular to the interface. 
 \item The strain and shear deformations are negligible, in such a way that layer and substrate are within the elastic limit at all times. 
 \item Substrate has no constraints to bend in neither of the two directions. This condition is no fully satisfied in the described experimental setup. As we use lever shaped substrates with one ends fixed to the holder, we constrain the bending along the short side. If the lever satisfies b $>$ 3a, with a and b the dimensions of the long and short sides, respectively, then the deformation in the transverse direction will not influence the bending along the long axis and this condition can be fulfil. 
\end{enumerate}

On the other hand, the crystal structure of the materials used in this work do not allow to fulfil the condition of isotropy. Eq.~\ref{eq:stoney} is not valid and it must be adjusted to the experimental conditions, taking into account the crystal orientation of the interface and the elastic constants of the material in the direction along which the curvature is measure. In this way, the biaxial modulus becomes:
\begin{equation}
\dfrac{1-\upsilon_S}{Y_s}\equiv M_S = c_{11}+c_{12}+2\dfrac{c_{12}^2}{c_{11}}
\label{eq:biaxial}
\end{equation}
where c$_{ij}$ are the substrate elastic constants. Rearranging Eq.~\ref{eq:stoney} for uniaxial stress along the [110] and [1-10] directions:
\begin{equation}
\dfrac{1}{R} = -\dfrac{6\sigma h}{h_S^2}\dfrac{M_S+2c_{44}}{4M_Sc_{44}}
\label{eq:stoney2}
\end{equation}
The value of c$_{44}$ is approximately M$_S$/2 so the error introduced by using Eq. \ref{eq:stoney} instead of \ref{eq:stoney2} is less than 2\%.

Until now, we have considered the accumulated stress introduced by a layer in an static situation. However during a MBE growth the stress variation might be due to changes in the deposited layer thickness, changes in its stress or even the surface reconstruction. For this reason, if the thickness of a layer changes $dh$ in a time $t+dt$, using a differential form of Eq.~\ref{eq:stoney}:
\begin{equation}
\dfrac{M_S h_S^2}{6}\dfrac{d\left(1/R\right)}{dt} = \sigma (z=h, t)\dfrac{dh}{dt}+\int _0^h\dfrac{d\sigma}{dt}dz+\left[\Delta \tau _S \right] 
\label{eq:diff}
\end{equation}
The right hand side of this equation have three terms. The first one describes changes in stress associated to an increase of thickness $h$ in a time interval [$t$, $t+dt$]. The second term accounts for relaxation processes in the already deposited layer at the time $t$. Finally, the third term is related with changes in the surface stress. We can define the accumulated stress at the time $t$ as:
\begin{equation}
\Sigma\sigma[h(t)] = \sigma (z=h, t)\dfrac{dh}{dt}+\int _0^h\dfrac{d\sigma}{dt}dz =  \int _0^{h(z)}\sigma(z)dz
\label{eq:accumulated}
\end{equation}
Substituting Eq. \ref{eq:accumulated} in \ref{eq:diff} we obtain:
\begin{equation}
\dfrac{M_S h_S^2}{6}\dfrac{d\left(1/R\right)}{dt} = \Sigma\sigma[h(t)]+\left[\Delta \tau _S \right] 
\label{eq:stoney3}
\end{equation}
As it can be seen, the magnitude measure in this experiments is the sum of the accumulated stress and the stress associated to changes in the surface reconstruction. Finally, combining Eq.~\ref{eq:curvatura} in its differential form (taking $\cos \alpha$ = 1), and Eq.~\ref{eq:stoney3} we get:
\begin{equation}
\Sigma\sigma[h(t)]+\left[\Delta \tau _S \right] = \dfrac{M_S h_S^2}{12}\dfrac{[d(t)-d_0]}{d_HL}
\label{eq:stress}
\end{equation}

\section{Implementation and characteristics of the technique}

An important improvement of this technique, as it is implemented in the Instituto de Microelectr\'onica de Madrid (IMM), is the use of a large area CCD camera to record the two beams simultaneously. This method has several advantages over the segmented detectors. On the one hand, the optical alignment is considerably easier since there is only one detector to be put in place to record both beams. On the other hand, it has larger dynamical range, as the reflected spots are recorded at all times regardless of their separation and exact positions (within a reasonable range). Finally, it has comparable resolution to the segmented detectors method without the need of low noise amplifiers or other extra equipment. 

%

Figure \ref{fig:setup} shows a detailed schema of the AS measurement system available at the IMM. The laser source (608 nm) produces an intense beam that hits a beam splitter, leading to two perfectly parallel beams of similar intensity. The beams cross the optical window of the MBE reactor and reach the sample perpendicularly to its surface. One of the beam, hereafter reference beam (RB), hits the fix end of the sample so its reflection is not affected by the growth process. The other beam, hereafter signal beam (SB), hits the free end of the sample and its reflection will be affected by the bending of the sample and hence by the stress accumulated during growth. The measurement of the reflected beams is performed in a backscattering geometry, minimizing the error introduced by the approximation made in Eq.~\ref{eq:stress}. 

The RB reaches directly the CCD whereas the SB crosses a prism to change its trajectory and send it to the camera. This prism is of capital importance in the setup and allows the usage of a CCD camera instead of the segmented detectors. Even in the case of the beams been reflected perfectly parallel, the distances between the spots would be of around 1 cm. In a more realistic case, where the beams diverge due to the deflection of the lever, the separation at a reasonable distance from the sample surface ($\sim$1 m in our case) could be of several cm, too large for most CCDs. Since the accumulated stress measurements depends only in the distance difference between the spots and not on their absolute value, this approach does not have any effect in the results.

We use a SpotOn CCD camera of Duma Optronics Ltd. and their acquisition software to get a beam positioning with sub-micron resolution. The distance between spots is sent to a custom software that converts it into accumulated stress, in real time, by means of Eq.~\ref{eq:stress}. This software also records the opening and closing of the effusion cell shutters, giving an exact match between the accumulated stress evolution and the materials growth. With this information, and assuming a typical distance of L = 975 mm between sample and detector, $d_H$ = 8 mm as the initial separation of the laser beams, and using the parameters characteristic of our substrates (GaAs, thickness $h_S$=100 $\mu$m, $M_S$ = 124 GPa), we obtain a maximum resolution of 0.02 N/m. This high resolution is normally not attainable due to vibrations and noise in the environment. Mechanical vacuum pumps, either from the MBE reactor or from nearby equipments, have the most detrimental effect and must be disconnected in order to perform high quality measurements. The real resolution in our system is normally between 0.05 and 0.1 N/m. 
 
\begin{figure}
	\centering
	\includegraphics{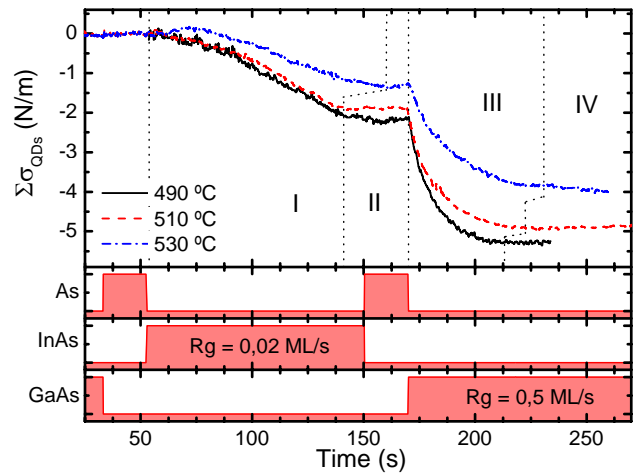}
	\caption{Stress accumulated during the formation of QDs after 2 ML of InAs. The lower part of the figure shows the materials that are growing at each time.}
	\label{fig:StressQDs}
\end{figure}

\section{Experimental results}
We have used this technique to characterize the growth of strain balanced quantum dots and quantum posts (QPs). This kind of nanostructures have a three dimensional shape and, thus, the common equations used to calculate the optimum strain balanced condition can not be used, as the strain is inhomogeneous and the layer thicknesses are not well defined.~\cite{Tatebayashi2009, Bailey2009, Alonso-Alvarez2010} In this work we use the in-situ accumulated stress measurements, as described above, to obtain the real stress that introduces the QDs and the most appropriate GaAsP thickness and composition that exactly compensates that stress. 

All samples have been growth using solid source MBE on GaAs (001) substrates 100 $\mu$m thick. InAs and GaAs/GaAsP growth rates are 0.02 and 0.5 ML/s, respectively. As beam equivalent pressure (BEP) is kept at 1.5$\times$10$^-6$ mbar at all times. 

\subsection{Strain balanced InAs quantum dots}

\begin{figure}
	\centering
	\includegraphics{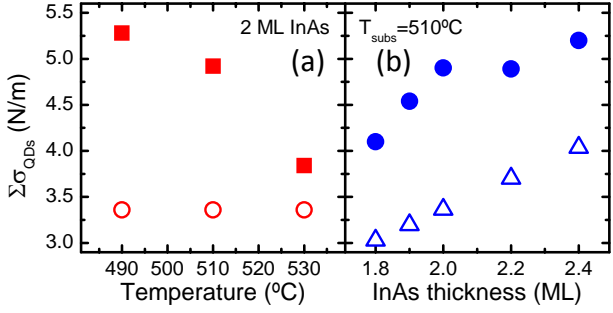}
	\caption{Accumulated stress per QD as a function of the growth temperature (a) and InAs thickness (b). Filled symbols represent the experimental data and empty ones the result of applying Eq.~\ref{eq:stressperlayer}}
	\label{fig:StressQDs2}
\end{figure}

In Fig.~\ref{fig:StressQDs} we show the evolution of the total accumulated stress as we grow an InAs QD layer for different substrate temperatures. As it was found by Silveira \textit{et al}, four regions can be distinguished:~\cite{Silveira2001} Region I: InAs begins to grow layer by layer, increasing the compressive stress linearly (except for a transition region at the beginning); Region II: just at the critical thickness, surface relaxes and QDs nucleate. The remaining deposited In keep floating on the surface or incorporates to the existing islands but without increasing the stress; Region III: during capping this remaining In incorporates, suddenly increasing the accumulated stress; And Region IV: when In is exhausted, GaAs grows without any further change in the stress. As it can be seen, the maximum accumulated stress depends strongly on the substrate temperature and also on the total amount of In deposited, as shown in Fig.~\ref{fig:StressQDs2}(1) and (b) (filled symbols). This kind of dependence is disregarded in the strain balance criteria used in QWs. The accumulated stress introduced by a flat, strained layer, assumed homogeneous, can be approximated by:
\begin{equation}
\Sigma\sigma_L = M_L \epsilon_L t_L
\label{eq:stressperlayer}
\end{equation}
where M$_L$, $\epsilon_L=(a_{subs}-a_L)/a_{subs}$ and t$_L$ are the layer biaxial modulus, the lattice mismatch between the layer and the substrate and the layer thickness respectively. Fig.~\ref{fig:StressQDs2} shows also the results of this equation applied to the nominal InAs thickness used in each case (open symbols). It can be seen that using the above equation to calculate the stress introduced by the QDs and, hence, the strain balanced condition, leads to sub-estimate the accumulated stress in all cases.
 
\begin{figure}
	\centering
	\includegraphics{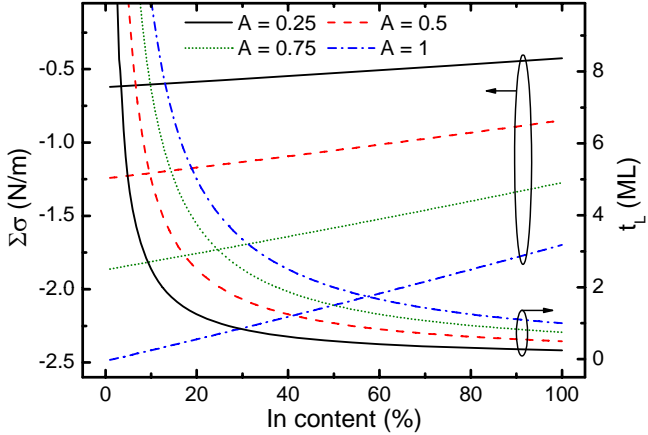}
	\caption{Evolution of the accumulated stress as a function of the In composition for a given total In amount. Right scale shows the corresponding layer thickness.}
	\label{fig:InComp}
\end{figure}

The reason for this discrepancy lays on the assumption that all the deposited In incorporates in the form of InAs. It is well known that, during the QDs capping, there is a large Ga-In intermixing, leading to quantum dots, wetting layer and capping made of InGaAs of varying composition. If the total In incorporated into the sample is to be constant, an InGaAs layer with a dilute alloy introduces more stress that a pure InAs layer. Fig.~\ref{fig:InComp} shows the dependence of the accumulated stress on the In content of the layer (x), keeping the restriction of equal overall In content:
\begin{equation}
	t_{L}\times x=A=constant \Longrightarrow \Sigma\sigma_L = M_L(x) \epsilon_L(x) a_{L}(x)\frac{A}{x}
	\label{eq:InComp}
\end{equation}
where $t_{L}$ is the layer thickness in ML. M$_L$(x) and a$_L$(x) are obtained as a linear interpolation of the GaAs and InAs parameters. As an example, if the In contained in a pure InAs monolayer is spread in two monolayers, giving a In$_{0.5}$Ga$_{0.5}$As alloy, the resulting accumulated stress changes from -1.7 N/m to -2.1 N/m. 

\begin{figure}
	\centering
	\includegraphics{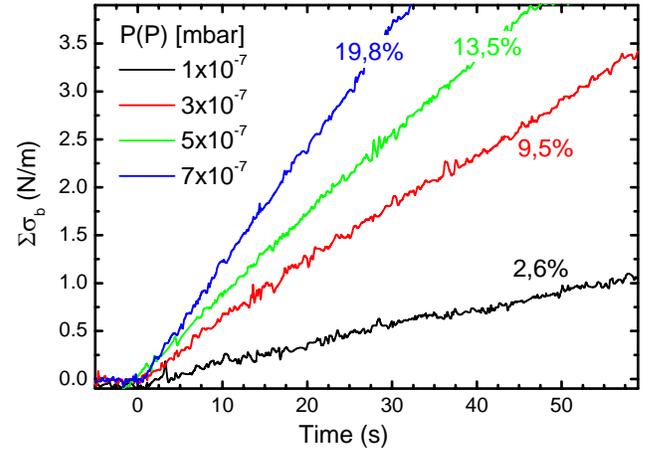}
	\caption{Accumulated stress as a function of the P BEP. Next to each curve is the P content of the alloy as estimated from Eq.~\ref{eq:stressperlayer}}
	\label{fig:StressGaAsP}
\end{figure}

A similar analysis can be performed for the growth of the GaAsP compensating layer. In this case, the variable parameter is the P BEP, which gives the GaAsP composition. Fig.~\ref{fig:StressGaAsP} shows the evolution of the accumulated stress as a function of time. As it is expected, the accumulated stress in this case is tensile, owing that the lattice parameter of GaAsP is smaller than that of GaAs. The composition of the layer can be estimated using equation \ref{eq:stressperlayer}, although it is not really needed for the calculation of the optimum strain balance condition. 

\begin{figure}
	\centering
	\includegraphics{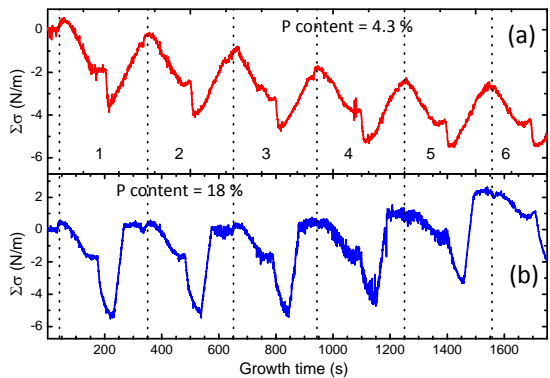}
	\caption{Accumulated stress of samples A (A) and B (b). On the left there is an schema of the samples layer structure.}
	\label{fig:Stack}
\end{figure}

Knowing the compressive stress introduced by the QDs and the tensile stress compensated by the GaAsP strain balanced layer (SBL) as a function of its composition, we designed two strain balanced QDs stacks (A and B) aiming to a 100\% of strain compensation. In both cases we use 2 ML of InAs for the QDs and a total spacer between layers of 15 nm. Substrate temperature and As BEP are kept constant at 510$\,^{\circ}\text{C}$ and 1.5$\times$10$^-6$ mbar respectively during the growth of the stacks. The substrate is GaAs (001) cantilever shaped (4x20 mm) with a thickness of 100 $\mu$m. The only difference between the samples is the compensating layer thickness and composition. In sample A we use a 13 nm thick SBL with 4.3\% of P after 1 nm of GaAs capping, whereas in sample B we use a SBL 5 nm thick and 18\% of P after 8 nm of GaAs capping. The evolution of stress during the growth of both stacks can be seen in Fig.~\ref{fig:Stack}(a) and (b).

Several things can be observed in these curves. Firstly, the average strain has been successfully balanced in both cases. Assuming that each QD layer introduces a stress of 5 N/m, we obtain an average strain compensation of 95\% for the sample A and 105\% for sample B. Secondly, in the sample A, the oscillations corresponding to the accumulation/compensation sequence are damped. This is due to the InAs and GaAsP intermixing and was expected given the small GaAs capping on top of QDs. The formation of quaternary InGaAsP compounds introduces a stress (compressive or tensile) smaller than the corresponding InAs or GaAsP alloys separately. Although the strain is balanced on average, the
stoichiometry of the stack is uncontrolled. This intermixing has an impact on the optical properties of the QDs and must be taken into account when placing the barrier too close to the QDs.

It should be notice that the results presented here are only an example of perfectly balanced stacks using the in-situ accumulated stress measurements. Other combinations of SBL composition and thickness are possible, such as using pure GaP layers, GaInP or dilute nitrides, having optical or electrical properties more suitable for a particular application. 

\subsection{Strain balanced quantum posts}

Quantum posts (QPs) are assembled by epitaxial growth of closely spaced quantum dot layers, modulating the composition of a semiconductor alloy, typically InGaAs. Contrary to normal self-assembled nanostructures, the height of the QPs can be controlled by the number of periods of the superlattice grown on top of the seed QDs layer. The amount of In in this kind of nanostructures is very large compared to stacked QDs, with the result that the accumulated stress is enormous and there is a tendency to the formation of dislocations. The largest QPs reported are about 40 nm high (Pendiente de revisar).

In this work we monitor the evolution of the accumulated stress in two QPs samples. The first one (sample C) uses a superlattice of 2.2 \AA\ of InAS and 8.5 \AA\ of GaAS grown at 510 $\,^{\circ}\text{C}$. In sample D, on the other hand, we substitute the GaAs by GaAsP with a nominal 14\% of P content. This approach has allowed us to fabricate extremely large QPs of up to 120 nm with very interesting optical and electronic properties, as it has been reported elsewhere.~\cite{Alonso-Alvarez2011a}

\begin{figure}
	\centering
	\includegraphics{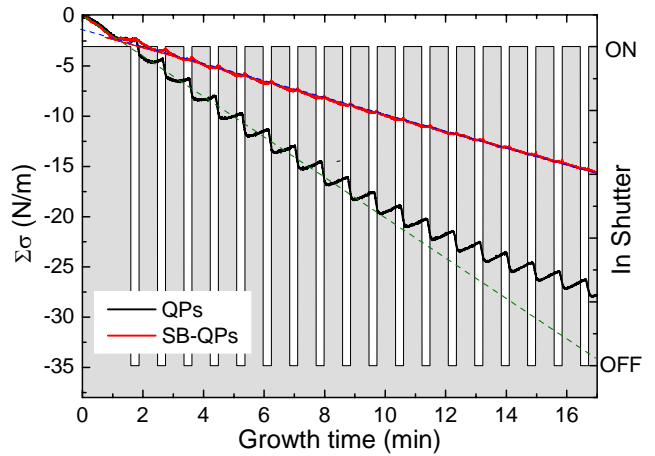}
	\caption{Accumulated stress of regular and strain balanced quantum posts (SB-QPs). Dashed lines are linear extrapolations of the accumulated stress in the first five periods in each case.}
	\label{fig:StressQPs}
\end{figure}

Fig.~\ref{fig:StressQPs} shows the resulting accumulated stress curves for both samples. Only the first 17 periods are measured since the sample bending became to large to be recorded with the CCD camera. For each curve, there is an approximated linear extrapolation of the stress accumulated by the first five periods, marked with dashed lines. Grey areas in the background represents the periods when the In effusion cell is open. The intermediate white regions represents a 10 s growth interruption under As flux plus the GaAs (GaAsP) growth.

As it can be seen, sample C accumulates larger stress than sample D. The degree of compensation in the superlattice (disregarding the QDs seed) can be estimated from the linear extrapolations mentioned above and gives a value of 57\%. The evolution of the accumulated stress in sample D is linear, behaviour that might be expected if all periods introduce the same amount of stress. The strain balanced oscillations are barely visible. As observed previously for QDs, this strong damping is directly related with the intermixing of the constituent materials. In sample C, it is remarkable the progressive bending of the accumulated stress that deviates from the linear tendency observed in sample D. Moreover, as it can be seen there is an apparent inversion of the accumulated stress during the In growth. In this periods, growing In reduces the accumulated stress, rather than increasing it. 

Both effects might be explained in terms of an initial stage of relaxation processes in the superlattice. As shown by Uj\'ue et al. during the growth of thick In$_{0.2}$Ga$_{0.8}$As layers on GaAs, prior to the formation of dislocations, there is an initial relaxation stage consisting on a ripening of the growth front along the [1$\overline 1$0] direction. From the point of view of accumulated stress, this effect produces a progressive deviation of the linear behaviour stated in Eq.~\ref{eq:stressperlayer}. The onset of the relaxation depends on the growth rate, taking place earlier for slow growth rates. This is roughly the situation found in sample C, where the average composition of In is also around 20\% and with an average growth rate of 0.07 ML/s.   

On the other hand, the reduction of the accumulated stress during In growth could be related also with a relaxation process but of a more local nature. The fabrication of QPs relays on the effective migration of In adatoms towards the top of buried QDs, where their elastic energy is smaller. The accumulation of this In atoms could partly relieve the stress of the InAs beneath them by locally increasing the lattice parameter of the structure. The growth of the GaAs capping suppresses this effect, increasing the stress by incorporating the InAs on the surface to the GaAs lattice structure.

The comparison of sample C and D also indicates that using a strain balanced technique to grow QPs is not only necessary in the case of extremely large QPs, but that it could be desirable also in average size nanostructures, with more than 8 or 10 periods, to avoid the relaxation processes described above. 

\section{Conclusions}

In this work we have shown the implementation of a very compact in-situ accumulated stress measurement setup based on the usage of a high resolution CCD camera to monitor the bending of the substrate during growth. The system outperforms previous designs in resolution and simplicity. We have used this system to study the strain balanced process of QDs and QPs. We have found that it is possible to achieve perfect strain compensation in QDs stacks by calibrating separately the stress introduced by the QDs and the compensating layer. This process depends strongly on the substrate temperature and the incorporation of In atoms to the sample. Finally, we have shown a reduction of 57\% in the stress accumulated during the growth of QPs by incorporating P to the matrix. The strain balanced technique is found to suppress the relaxation processes that take place in the first stages of the grow of this nanostructures. In summary, this experiments show the capability of strain balance technique to improve the quality of quantum nanostructures and the importance of kinematics in the optimization of the optimum strain balanced conditions.

\section{Acknowledgements}

We acknowledge the financial support by MICINN (TEC2008-06756-C03-01/03, ENE2009-14481-C02-02, CSD2006-0004, CSD2006-0019), CAM (S2009ESP-1503, S2009/ENE-1477) and CSIC (PIF 200950I154). 

\bibliography{papers3}
\end{document}